\documentclass{article}
\usepackage{amsmath}
\usepackage{amssymb}

\newtheorem{theorem}{Theorem}

\newtheorem{cor}[theorem]{Corollary}
\newtheorem{pro}[theorem]{Proposition}
\newtheorem{df}[theorem]{Definition}
\newtheorem{rmk}[theorem]{Remark}

\newcommand\calN{{\cal N}}

\newcommand\bfZ{\bf Z}
\newcommand\bfC{\bf C}

\newcommand\dprime{\prime\prime}
\newcommand\evl{\alpha_t}

\newcommand{\norm}[1]{\left\Vert#1\right\Vert}

\begin{document}
\newpage\thispagestyle{empty}
\begin{center}
{\Huge\bf
On Spectral Gap,  
\\
U(1) Symmetry 
\\
and
\\
Split Property
\\
in
\\
 Quantum Spin Chains }
\\
\bigskip\bigskip
\bigskip\bigskip
{\Large Taku Matsui}
\\
 Graduate School of Mathematics, Kyushu University,
\\
1-10-6 Hakozaki, Fukuoka 812-8581, JAPAN
\\
 matsui@math.kyushu-u.ac.jp
\\
\bigskip\bigskip
August, 2008
\end{center}
\bigskip\bigskip\bigskip\bigskip
\bigskip\bigskip\bigskip\bigskip
{\bf Abstract:}  In this article,  we consider a class of ground states for quantum spin chains 
on an integer lattice. First we show that presence of the spectral gap between the ground state
energy and the rest of spectrum implies the split property of certain subsystems.  
As a corollary, we show that gapless excitation exists  for spinless Fermion 
if the pure gauge invariant ground state is non-trivial  and translationally invariant.
\\
\\
{\bf Keywords:} quantum spin chain, spectral gap, $U(1)$ gauge symmetry,
Haag duality, Bell's inequality,Lieb-Robinson bound.
\\
{\bf AMS subject classification:} 82B10 

\newpage
\section{Introduction.}\label{Intro}
\setcounter{theorem}{0}
\setcounter{equation}{0}
In what follows, we show presence of the spectral gap between the ground state
energy and the rest of spectrum implies statistical independence of 
left and right semi-infinite subsystems. This independence is called {\it split property} 
in the context of local quantum field theory (local QFT) in th sense of R. Haag.
(c.f. \cite{Haag})  
The split property is known to hold in a number of situations of local QFT
while, as far as we are aware,  this condition  for quantum spin chains was never proved
in the situation we consider here.
As a corollary we will see that
a gapless excitation exists in certain $U(1)$ symmetric quantum lattice models on $\bfZ$ .   
\\
Now we will exhibit typical examples of Hamiltonians we have in our mind.
These are $U(1)$ gauge invariant  finite range
Hamiltonians for quantum spin chains such as the Heisenberg Hamiltonians $H_{XXX}$
on the one-dimensional integer lattice $\bfZ$  or fermionic systems on $\bfZ$, 
$H_F$ as described as follows:
\begin{equation}
H_{XXX}=  \sum_{i,j \in \bfZ}  \sum_{\alpha =x,y,z}  J_{ij} \sigma_{\alpha}^{(i)} \sigma_{\alpha}^{(j)},
\label{eqn:x1}
\end{equation}
\begin{equation}
H_{F} = \sum_{i,j \in \bfZ} t_{ij} c^{*}c_{j} +  \sum_{k \in \bfZ}  V_{k} (n),
\label{eqn:x2}
\end{equation}
where $\sigma_{\alpha}^{(j)}$ in (\ref{eqn:x1}) is the spin operator at the site $j$
 in which the direction of the spin is denoted by $\alpha$.  
 $c^{*}_{j}$ ,$c_{i}$ are Fermion creation - annihilation operators satisfying the anti-commutation relations and  
 $J_{ij}$ and $ t_{ij}$ are coupling constants satisfying the conditions
 (finite rangeness and translational invariance);
$$ J_{ij} = J_{i-j 0}, \: t_{ij} = t_{i-j 0}, \quad  J_{i 0}=0, \: t_{i 0}=0 $$
if $|i| >r$. 
$V_{k} (n)$ is a polynomial of the local number operators  $n_{i} = c^{*}_{i}c_{i}$
at the site $i$ . 
A simple example of $V_{k} (n)$ is $V_{k} (n) = K n_{k}n_{k+1}$ ($K$: a constant).
\par
Both $H_{XXX} $ and $H_F$ are invariant under the global $U(1) $ gauge transformation
where, for the spin chain, the gauge transformation is defined via the rotation around 
the z axis and, for Fermion system, it is defined via the formula
$c_{j} \to e^{i\theta} c_{j}$. Then, one of our results is expressed as follows.
\begin{theorem}
\label{th:main}
\noindent
\newline
(i) 
 Consider the quantum spin chain on $\bfZ$ and the spin at each site is $1/2$.
Let $H_{S}$ be a translationally invariant , $U(1)$ gauge invariant  finite range Hamiltonian. 
Suppose that $\varphi$ is a $U(1)$ gauge invariant , translationally invariant pure ground state
of $H_{S}$.
If $\varphi$ is not a product state, a  gapless excitation exists between the ground state energy
and the rest of the spectrum of the effective Hamiltonian on the GNS representation space.
 \\
 (ii) Consider the spinless Fermion lattice system on  $\bfZ$.
Let $H_{F}$ be a translationally invariant , $U(1)$ gauge invariant  finite range Hamiltonian. 
Suppose that $\varphi$ is a $U(1)$ gauge invariant , translationally invariant pure ground state
of $H_{F}$.
Suppose further that $\varphi$ is not neither  the standard Fock state  $\psi_{F}$
nor the standard anti-Fock state, a gapless excitation exists between the ground state energy
and the rest of the spectrum of the effective Hamiltonian on the GNS representation space .
\end{theorem}
In the above theorem, by the standard Fock state we mean the state $\psi_{F}$
 specified by the identity $\psi_{F}(c^{*}_{j}c_{j})=0$ for any $j$ and
  the standard anti-Fock state is the state $\psi_{AF}$ specified by the identity  
  $\psi_{AF}(c_{j}c^{*}_{j})= 1$  for any $j$.
\\
\\
The infinite volume ground state we consider here is defined in \cite{BratteliRobinsonII}.
$\varphi$ is a ground state for an infinite volume Hamiltonian $H$ if $\varphi$
is a normalized positive functional on the $C^*$-algebra of quasi-local observables satisfying
\begin{equation}
 \varphi (Q^{*}[ H , Q] ) \geq 0
\label{eqn:x3}
\end{equation}
for any  quasi-local observable $Q$.
The infinite volume limit of ground states for finite volume
Hamiltonians with any boundary conditions gives rise to a state satisfying
(\ref{eqn:x3}). More precisely let $H_{n}$ be a sequence of finite volume Hamiltonians
satisfying $\lim [H_{n}, Q] = [H,Q]$ for any local observable $Q$
and $\Omega_{n}$ be a unit eigenvector for the least
eigenvalue of $H_{n}$. Set  
$$\varphi (Q) =\lim_{n} (\Omega_{n} , Q \Omega_{n}) .$$
Then, the state $\varphi$ satisfies the inequality (\ref{eqn:x3}).
\bigskip
\noindent
\par
Results similar to Theorem\ref{th:main} were obtained before for several cases.
For example, for antiferromagnetic Heisenberg model, presence of gapless excitation 
was proved by I.Affleck and E.Lieb in \cite{AffleckLieb} .  
Results on Fermion models were obtained by T.Koma in \cite{Koma2007} .
The difference between previous results and ours lies in the two points.
First our result is on ground state for arbitrary boundary conditions.
The second point is that our proof is new and the argument is based on
three mathematical ingredient: 
\\
(1)  Results on Bell's inequality  for infinite quantum systems due to
S.Summers and R.F.Werner (\cite{SummersWerner}) , 
\\
(2) Haag duality of quantum spin chain recently proved by us \cite{KMSW2} and 
\\
(3) Improved Lieb-Robinson bound due to R.Simms and B.Nachteregaele 
\cite{NachtergaeleSims2007Dec}.
\\
\\
We will see that these three results imply that any translationally invariant pure  ground states
have the split property if there is a spectral gap between the ground state energy and the rest
of spectrum. More precisely, we will prove the following theorem.
\begin{theorem}
\label{th:mainSplit}
\noindent
 \newline
 (i) Consider a quantum spin chain on $\bfZ$.
Let $H_{S}$ be a translationally invariant  finite range Hamiltonian and let $\varphi$ be
a translationally invariant pure ground state of $H_{S}$.
Suppose that  there is a  gap between the ground state energy
and the rest of the spectrum of the effective Hamiltonian on the GNS representation space
associated with  $\varphi$ .
\par
Then,
$\varphi$ is equivalent to a product state $\psi_{L}\otimes \psi_{R}$ 
where  $\psi_{L}$ is a state of the algebra of observables localized in $(-\infty, 0]$
and $\psi_{R}$ is a state of the algebra of observables localized in $[1, \infty)$
 \\
 (ii) Consider a Fermion lattice system on  $\bfZ$ with a finite number of components
 at each lattice site.
Let $H_{F}$ be a translationally invariant  finite range Hamiltonian and
let $\varphi$ be a translationally invariant pure ground state.
Suppose that  there is a  gap between the ground state energy
and the rest of the spectrum of the effective Hamiltonian on the GNS representation space
associated with  $\varphi$ .
Then,  $\varphi$ is equivalent to a product state $\psi_{L}\otimes_{Z_{2}} \psi_{R}$ 
where  $\psi_{L}$ is a even state of the algebra of observables localized in $(-\infty, 0]$
and $\psi_{R}$ is a even state of the algebra of observables localized in $[1, \infty)$
and by $\otimes_{Z_{2}}$ we denote the graded tensor product.
\end{theorem}
The notion of {\it graded tensor product}  is fermion analogue of the tensor product
which we introduce in Section 4.

We employ the $C^{*}$-algebraic method to prove Theorem \ref{th:main} and 
Theorem \ref{th:mainSplit}.
The method is an abstract functional analysis which can be applied to Hamiltonians
with a form more general  than in (\ref{eqn:x1}) and (\ref{eqn:x2}) .
The standard references for the framework and basic notions of  the $C^{*}$-algebraic method 
are \cite{BratteliRobinsonI} and \cite{BratteliRobinsonII}.  
In Section 2, we introduce several notions to describe our results precisely.
We introduce  Lieb-Robinson bound and uniform exponential clustering.
In Section  3, we prove  twisted Haag duality for the fermionic system.
In section 4, we explain the reason why that the spectral gap implies split property 
with the aid of Bell's inequality in infinite quantum systems.
We present our proof of Theorem \ref{th:main} and Theorem \ref{th:mainSplit} in the final section.
\section{Infinite Volume Ground States and Spectral Gap.}\label{Cstar}
\setcounter{theorem}{0}
\setcounter{equation}{0}
First we introduce several notations and notions of quantum spin chain on $\bfZ$ and 
then later we mention the case of Fermions.
We denote the $C^{*}$-algebra of (quasi)local observables by $\frak A$. 
This means that $\frak A$ is the UHF $C^*-$algebra $n^{\infty}$ 
( the $C^{*}$-algebraic completion of the infinite tensor product of n by n matrix algebras ):
$${\frak A} = \overline{\bigotimes_{\bfZ} \: M_{n}({\bf C})}^{C^*} , $$
where $M_{n}({\bf C})$ is the set of all n by n complex matirces.
Each component of the tensor product is specified with a lattice
site $j \in \bfZ$.  ${\frak A}$ is the totality of quasi-local observables.
We denote by $Q^{(j)}$ the element of ${\frak A}$ with $Q$ in
the j th component of the tensor product and the identity in any other
components :
$$  Q^{(j)} = \cdots \otimes 1 \otimes 1 \otimes Q
\otimes 1 \otimes 1 \otimes \cdots$$
For a subset $\Lambda$ of $\bfZ$ , ${\frak A}_{\Lambda}$ is defined as
the $C^*$-subalgebra of ${\frak A}$ generated by elements $Q^{(j)}$ 
with all $j$ in $\Lambda$.
We set
$${\frak A}_{loc} = \cup_{ \Lambda \subset {\bfZ} : | \Lambda | < \infty}
 \:\: {\frak A}_{\Lambda}  $$
where the cardinality of $\Lambda$ is denoted by $|\Lambda |$.
We call an element of ${\frak A}_{loc} $ a local observable
or a strictly local observable. 
\par
By a state $\varphi$ of a quantum spin chain, we mean a normalized positive linear functional
on  $\frak A$ which gives rise to the expectation value of a quantum mechanical state.
\par
When $\varphi$ is a state of ${\frak A}$, the restriction of $\varphi$
to ${\frak A}_{\Lambda}$ will be denoted by $\varphi_{\Lambda}$ :
$$\varphi_{\Lambda} = \varphi \vert_{{\frak A}_{\Lambda}} .$$
We set 
$${\frak A}_R =  {{\frak A}}_{[1,\infty)} \: , \:
{\frak A}_L =  {\frak A}_{(-\infty, 0]} \: , \:  \varphi_R=\varphi_{[1,\infty)} \: , \:
\varphi_L = \varphi_{(-\infty, 0]}  \: \: .$$
By $\tau_j$, we denote the automorphism
of $\frak A$ determined by 
$$ \tau_j(Q^{(k)})=Q^{(j+k)}$$ 
for any j and k in $\bfZ$. $\tau_j$ is referred to as the lattice translation of ${\frak A}$.
\par
Given a state $\varphi$ of $\frak A$, we denote the GNS representation of $\frak A$ associated with
$\varphi$ by $\{ \pi_{\varphi}(\frak A) , \Omega_{\varphi}, \frak H_{\varphi} \}$ where
 $\pi_{\varphi}(\cdot)$ is the representation of $\frak A$ on the GNS Hilbert space $ \frak H_{\varphi} $ 
 and $\Omega_{\varphi}$ is the GNS cyclic vector satisfying
 $$\varphi( Q) = \left( \Omega_{\varphi},  \pi_{\varphi}(Q)\Omega_{\varphi}\right) 
 \quad  \: Q \in \frak A .$$
 Let $\pi$ be a representation of $\frak A$ on a Hilbert space.
The von Neumann algebra generated by $\pi({\frak A}_{\Lambda})$
is denoted by ${\frak M}_{\Lambda}$. We set
$${\frak M}_R = {\frak M}_{[1,\infty)} =\pi ({\frak A}_{R})^{\dprime} ,  \quad
{\frak M}_L = {\frak M}_{(-\infty , 0]} =\pi ({\frak A}_{L})^{\dprime} .$$
\bigskip
\par
In terms of the above definitions, we introduce the notion of the dynamics 
(the Heisenberg time evolution) and the ground state for infinite volume systems.
By {\it Interaction} we mean an assignment $\{\Psi (X) \}$ of each finite subset $X$ of 
$\bfZ$ to a selfadjoint operator $\Psi (X)$ in $\frak A_{X}$.
We say that an interaction is of finite range if there exists a positive number $r$ such
that    $\Psi (X) = 0$ if that  the diameter of $X$ is larger than  $r$.
An interaction is translationally invariant if and only if 
$ \tau_{j}(\Psi (X)) = \Psi (X+j)$ for any $X \subset \bfZ$ and for any $j \in \bfZ $.
For a translationally invariant finite range interaction,
 we consider the formal infinite volume Hamiltonian $H$ which
 is an infinite sum of local observables.
$$H =   \sum_{X \subset \bfZ} \Psi (X).$$ 
This sum does not converge in the norm topology, however
the following commutator  and the limit make sense:
$$  [ H ,  Q ]  = \lim_{n \to\infty} [ H_{n} , Q] = \sum_{X \subset \bfZ} [ \Psi (X) , Q ]   , \:\:
 \lim_{n \to\infty} e^{it H_{n}} Q  e^{-it H_{n}}   \:\:  Q \in \frak A_{loc} $$
where 
$ H_{n} =  \sum_{X \subset [-n,n]} \Psi (X) $.
\par
More generally, for any finite subset $\Lambda$ the finite volume Hamiltonian $H_{\Lambda}$
 is defined by $H_{\Lambda} =  \sum_{X \subset \Lambda} \Psi (X)$.
Then if  $\{ \Psi (X) \}$ is a translationally invariant interaction, and  if we assume that 
\begin{equation}
\sum_{X \ni 0} |X| || \Psi (X) || < \infty , 
\label{eqn:y1}
\end{equation}
the following limit exists
$$\evl (Q) =  \lim_{ \Lambda \to \bfZ} e^{itH_{\Lambda}}  Q e^{-itH_{\Lambda}}$$  
for any element $Q$ of $\frak A$ in the $C^{*}$ norm topology. 
We call $\evl (Q)$ the time evolution of $Q$.
\begin{df}
Suppose the time evolution $\evl (Q)$ associated with an interaction
satisfying  (\ref{eqn:y1}) is given.
Let $\varphi$ be a state of $\frak A$. $\varphi$ is a ground state of $H$ if and only if
\begin{equation}
\varphi (Q^{*} [H , Q] )  =  \frac{1}{i} \frac{d}{dt} \varphi (Q^{*} \evl (Q)) \geq 0  
\label{eqn: y2}
\end{equation}
for any $Q$ in $\frak A_{loc}$.
\end{df}
Suppose that $\varphi$ is a ground state for $\evl$ . In the GNS representation of
\\
$\{ \pi_{\varphi}(\frak A) , \Omega_{\varphi}, \frak H_{\varphi} \}$, there exists a positive
selfadjoint operator $H_{\varphi} \geq 0$ such that 
$$ e^{itH_{\varphi}} \pi_{\varphi}(Q) e^{-itH_{\varphi}} = \pi_{\varphi}(\evl (Q)), \quad
 e^{itH_{\varphi}} \Omega_{\varphi} =  \Omega_{\varphi} $$
for any $Q$ in $\frak A$.
Roughly speaking, the operator $H_{\varphi}$ is the effective Hamiltonian
 on the physical Hilbert space $ \frak H_{\varphi}$ obtained after regularization
 via subtraction of the vacuum energy.
\\
\\
\par
The spectral gap we are interested in is that of $H_{\varphi}$.
Note that, in principle, a different choice of a ground state gives rise
to a different spectrum.
\begin{df}
We say that $H_{\varphi}$ has the spectral gap 
if $0$ is a non-degenerate eigenvalue of $H_{\varphi}$ and for a positive $M >0$,
$H_{\varphi}$ has no spectrum in $(0,M)$,i.e. $Spec (H_{\varphi}) \cap (0,M) = \emptyset $. 
\end{df}
It is easy to see that $H_{\varphi}$ has the spectral gap  if and only if there exists
 a positive constant $M$ such that 
$$\varphi (Q^{*} [H , Q] ) \geq M (\varphi(Q^{*}Q) - |\varphi (Q)|^{2} ) .$$
\\
\\
\par
In the quantum field theory with locality it is known that presence of the spectral
gap implies exponential decay of (spacial) correlation. (c.f.\cite{ArakiRuelle},
\cite{Fredenhagen}) The most general result is obtained by K.Fredenhagen
and the result is referred to as Fredenhagen's cluster theorem.
\par
The nature of locality in quantum spin chains is quite different from 
 the relativistic quantum field theory as we do not have speed of light.
 Nevertheless, there is a control of propagation of quasi locality 
 which is due to E.Lieb and D.Robinson.
 The bound of this kind is called {\it the Lieb-Robinson bound}.
(  \cite{LiebRobinson}) The inequality is  described as follows.
 \begin{equation}
 ||  [ \evl (Q) , R ] || \leq C(Q,R)  e^{-a d(X,Y)- v |t| } 
 \label{eqn:y3}
 \end{equation}
 where $Q$(resp. $R$) is an element of $\frak A_{X}$ (resp. $\frak A_{Y}$ and 
  $C(Q,R)$  is a constant positive depending on  $Q$ and $R$ and 
  $v$ is another positive constant called ``group velocity''.
Though not straightforward, once the above quasi-locality estimate is established
we may expect a lattice model analogue of  Fredenhagen's cluster theorem.
This was achieved relatively recently. 
( See \cite{nachtergaele2005}, \cite{Koma} and see also 
\cite{bravyi2006}, \cite{Koma2007},   \cite{NaOgSi},
 \cite{nasi} for extension and application of the results.)
The estimate of spacial decay of correlation
we need for our purpose is due to B.Nachtergaele and R.Sims.
Now we present this version in \cite{NachtergaeleSims2007Dec} .
\begin{theorem}[B.Nachtergaele and R.Sims 2007] 
We consider the quantum spin chain on $\bfZ$.
Assume that the interaction $\{\Psi (X)\}$ is translationally invariant and of finite range.
Let $\varphi$ be a translationally invariant pure ground state of the Hamiltonian $H$.
Assume further that the effective Hamiltonian $H_{\varphi}$ has the spectral gap.
\\
Then, there exists positive constants $C$ and $K$ such that 
for any $Q$ in $\frak A_{L}$, $R$ in $\frak A_{R}$ and any positive integer $j$,
 the following estimate is valid.
 \begin{equation}
| \varphi (Q \tau_{j}(R))  -  \varphi (Q )  \varphi (R)  | \leq C || Q||\cdot ||R|| e^{-K j}
 \label{eqn:y4}
 \end{equation}
$C$ and $K$ are independent of $Q$ and $R$. 
\label{th:Exponential}
\end{theorem}
That $C$ is independent on the size of support of observables $Q$ and $R$ is crucial 
to our argument below and it seems that this independence was never obtained before
 \cite{NachtergaeleSims2007Dec}.
\\
\\
Next we introduce the $U(1)$ gauge action to describe our main result (Theorem \ref{th:main})
 more precisely.
We consider the spin 1/2 system , thus $\frak A$ is the infinite tensor product of
2 by 2 matrix algebras. We set
\begin{equation}
S_{z} = \frac{1}{2} \sum_{j=-\infty}^{\infty} \sigma_{z}^{(j)}  , 
\:\: \gamma_{\theta} (Q)  =  e^{i\theta S_{z}} Q  e^{-i\theta S_{z}}
\label{eqn:y500}
\end{equation}
Then $\gamma_{2\pi}(Q) =Q$ for any $Q$ in $\frak A$ and $\gamma_{\theta} $ gives rise
a  $U(1)$ action on $\frak A$. 
Instead of proving Theorem \ref{th:main}, we will show an equivalent result stated as follows. 
\begin{theorem}
\label{th:main2}
Suppose that the spin $S$ is one half,  hence 
the one-site observable algebra is $M_{2}({\bf C})$. 
Let $\varphi$ be a translationally invariant pure ground state of
a finite range translationally invariant Hamiltonian.
Suppose further that the effective Hamiltonian $H_{\varphi}$ has the spectral gap
and $\varphi$ is $\gamma_{\theta} $ invariant for any $\theta$.
Then, $\varphi$ is a product state.
\end{theorem}
\bigskip
\noindent
\newline
Next we consider fermionic systems. The results discussed above are extended
in a natural way. Let $\frak A^{CAR}$ be the CAR
(canonical anti-commutation relations) algebra generated by Fermion creation-annihilation operators 
 $c_j$ and $c^*_k$ ($j,k \in \bfZ$) satisfying  
 \begin{equation}
\{ c_j , c_k \} =0,\:\: \{ c^*_j , c^*_k \} =0,\:\: \{ c_j , c^*_k \} = \delta_{jk} 1
\label{eqn:y5}
\end{equation}
For each subset $\Lambda$ of $\bfZ$, $\frak A^{CAR}_{\Lambda}$
is the $C^*$-subalgebra generated by $c_j$ and $c^*_k$ for $j,k \in \Lambda$.
$\frak A^{CAR}_{loc}$, $\frak A^{CAR}_{L}$, $\frak A^{CAR}_{R}$
are defined as before. The  {\em paritiy}  $\Theta$ 
is an automorphism of $\frak A^{CAR}$ defined by
 $\Theta (c_j ) = -c_j$ , $\Theta (c^*_j) = -c^*_j $ for any $j$. 
We set
$$ (\frak A^{CAR})_{\pm} = \{ Q \in \frak A^{CAR}  \: | \:  \Theta (Q) = \pm Q  \} ,$$
$$ (\frak A^{CAR}_{\Lambda})_{\pm} =  (\frak A^{CAR})_{\pm} \cap \frak A^{CAR}_{\Lambda},
\quad  (\frak A^{CAR}_{loc})_{\pm} =  (\frak A^{CAR})_{\pm} \cap \frak A^{CAR}_{loc}.$$
$\tau_{j}$ is the shift automorphisms defined by
$$\tau_{j}(c_{k}) = c_{k+j} , \quad \tau_{j}(c^{*}_{k}) = c^{*}_{k+j}.$$
We introduce the $U(1)$ gauge action $\gamma_{\theta}$ via the following equation:
 $$ \gamma_{\theta} (c_{j})  =  e^{-i\theta} c_{j} ,  \quad
  \gamma_{\theta} (c^{*}_{j})  =  e^{i\theta} c^{*}_{j} $$
For fermionic systems, an interaction is   an assignment $\{\Psi (X) \}$ of each finite subset $X$ of 
$\bfZ$ to a selfadjoint operator $\Psi (X)$ in $(\frak A^{CAR}_{X})_{+}$.
If we assume finite rangeness and translational invariance
of interactions and the formal infinite volume Hamiltonian
$H =   \sum_{X \subset \bfZ} \Psi (X)$ gives rise to the time evolution
$\evl$ of the system via the formula,
$$\frac{d}{dt} \evl (Q)|_{t=0} =  [ H ,  Q ] .$$ 
The notions of the effective Hamiltonian and the spectral gap are formulated as before
and the Lieb-Robinson bound is valid for $\Theta$-twisted commutators.
 \begin{equation}
 ||  \{ \evl (Q) , R \} || \leq C(Q,R)  e^{-a d(X,Y)- v |t| } 
 \label{eqn:y6}
 \end{equation}
 where $Q$(resp. $R$) is an element of $(\frak A^{CAR}_{X})_{-}$ (resp. $(\frak A^{CAR}_{Y})_{-}$). 
 \begin{theorem} 
We consider the spinless Fermion on $\bfZ$.
Assume that the interaction $\{\Psi (X)\}$ is translationally invariant and of finite range.
Let $\varphi$ be a translationally invariant pure ground state of the Hamiltonian $H$.
Assume further that the effective Hamiltonian $H_{\varphi}$ has the spectral gap.
\\
Then, there exists positive constants $C$ and $K$ such that 
for any $Q$ in $\frak A^{CAR}_{L}$, $R$ in $\frak A^{CAR}_{R}$ and any positive integer $j$,
 the following estimate is valid.
 \begin{equation}
| \varphi (Q \tau_{j}(R))  -  \varphi (Q )  \varphi (R)  | \leq C || Q||\cdot ||R|| e^{-K j}
 \label{eqn:y7}
 \end{equation}
 $C$ and $K$ are independent of $Q$ and $R$. 
\label{th:Exponential2}
\end{theorem}
The standard Fock state $\psi_{F}$ is a state of Fermion which is determined uniquely by the formula
 \begin{equation}
 \psi_{F} (c^{*}_{j} c_{j} ) = 0
 \label{eqn:y8}
 \end{equation}  
for any $j$. Similarly, the standard anti-Fock state
 $\psi_{AF}$ is a state of Fermion which is  determined uniquely by the formula
 \begin{equation}
  \psi_{AF} (c_{j} c^{*}_{j} ) = 0
 \label{eqn:y9}
 \end{equation}  
   for any $j$.
\begin{theorem}
\label{th:main3}
Consider the spinless Fermion on $\bfZ$ and  
let $\varphi$ be a translationally invariant pure ground state for
a finite range translationally invariant Hamiltonian.
Suppose further that the effective Hamiltonian $H_{\varphi}$ has the spectral gap
and $\varphi$ is $\gamma_{\theta} $ invariant for any $\theta$.
Then, $\varphi$ is either $\psi_{F}$  or  $\psi_{AF}$ .
\end{theorem}
\newpage
\section{Twisted Haag duality}\label{Haag}
\setcounter{theorem}{0}
\setcounter{equation}{0}
In this section, we show twisted Haag duality for translationally invariant 
pure states of Fermion systems.
First, let us recall the definition of the Haag duality for quantum spin chains.
Consider an irreducible representation $\pi (\frak A )$ of $\frak A$ on a Hilbert space $\frak H$.
We say Haag duality holds for a subset $\Lambda$ in $\bfZ$
if  $   \pi ({\frak A}_{\Lambda})^{\dprime}  = \pi ({\frak A}_{\Lambda^{c}})^{\prime}$ 
where $\Lambda^{c}$ is the complement of  $\Lambda$ in  $\bfZ$.
At first sight, this duality may be
expected. However, if one recalls other examples of infinite quantum systems
such as positive energy representation of loop groups,
 the duality turns out to be highly non-trivial. (c.f. \cite{Wassermann})
 In the loop group case when we choose the upper semi-circle  
 as $\Lambda$, the duality does not hold for non-vacuum sectors of positive energy representations.
 \\
 If $\pi (\frak A_{\Lambda})^{\dprime}$ is a type I von Neumann algebra, it is easy to show
 Haag duality for any  $\Lambda$ in $\bfZ$.  If $\Lambda$ is the semi-interval $[1, \infty)$
 and the representation is associated with a (gapless) ground state
 $\frak M_{R} =   \pi (\frak A_{\Lambda})^{\dprime}$ can be of non type I.
Nevertheless in \cite{KMSW2}, we succeeded in proving the Haag duality
for $\frak M_{R}$ in GNS representations of translationally invariant pure states
\begin{theorem}
Let $\varphi$ be a translationally invariant pure state of the UHF algebra
$\frak A$, and let $\{\pi_{\varphi} (\frak A), \Omega_{\varphi} , {\frak H}_{\varphi}  \}$ 
be the GNS triple for  $\varphi$.
Then, the Haag duality holds:
\begin{equation}
{\frak M}_R = {\frak M}_L^{\prime}
\label{eqn:z1}
\end{equation}
\label{th:dualitySpin}
\end{theorem}
Next we consider the GNS representation of $\frak A^{CAR}$
associated with a translationally invariant pure state $\psi$ and we show the fermionic
version of Haag duality.
In general, any translationally invariant factor state $\psi$ of  $\frak A^{CAR}$
is $\Theta$ invariant. (See \cite{ArakiMoriya} for proof.) 
Suppose that a state $\psi$ of  $\frak A^{CAR}$ is $\Theta$ invariant and 
let $\{\pi_{\psi} (\frak A_{CAR}), \Omega_{\psi}, \frak H_{\psi}\}$ be the GNS triple 
associated with $\psi$.
There exists a (unique) selfadjoint unitary $\Gamma$ on $\frak H_{\psi}$ 
satisfying
\begin{equation}
\Gamma \pi_{\psi}(Q)\Gamma^{-1}= \pi_{\psi}(\Theta(Q)) , \:\: 
\Gamma^{2}=1, \:\: \Gamma =\Gamma^{*}, \:\: \Gamma \Omega_{\psi} = \Omega_{\psi}. 
\label{eqn:z2}
\end{equation}
With aid of $\Gamma$, we introduce another representation $\overline{\pi}_{\psi}$
of $\frak A^{CAR}$ via the following equation:
\begin{equation}
\overline{\pi}_{\psi}(c_{j}) = \pi_{\psi} (c_{j})\Gamma, \:\: 
\overline{\pi}_{\psi}(c_{j}^{*}) = \Gamma \pi_{\psi} (c_{j}^{*})
\label{eqn:z3}
\end{equation}
for any integer $j$.
\par
Let $\Lambda$ be a subset of $\bfZ$ and $\psi$ be a state of  $\frak A^{CAR}$ which is $\Theta$ invariant. 
By definition,$ \pi_{\psi}( \frak A^{CAR}_{\Lambda} )^{\dprime} \subset
\overline{\pi}_{\psi} ( \frak A^{CAR}_{\Lambda^{c}} )^{\prime}  $.
 We say the twisted Haag duality is valid for  $\Lambda$ if and only if
\begin{equation}
\pi_{\psi}( \frak A^{CAR}_{\Lambda} )^{\dprime} = 
\overline{\pi}_{\psi} ( \frak A^{CAR}_{\Lambda^{c}} )^{\prime} 
\label{eqn:z4}
\end{equation}
holds.
\begin{theorem}
Let $\psi$ be a translationally invariant pure state of the CAR algebra
$\frak A^{CAR}$. and let $\{\pi_{\psi} (\frak A^{CAR}), \Omega_{\psi} , {\frak H}_{\psi}  \}$ 
be the GNS triple for  $\psi$.
Then, the twisted Haag duality holds for $\Lambda =[1,\infty )$.
\begin{equation}
\pi_{\psi}((\frak A^{CAR})_L)^{\dprime} = 
\overline{\pi}_{\psi} ((\frak A^{CAR})_{R} )^{\prime} 
\label{eqn:z5}
\end{equation}
\label{th:dualityFermi}
\end{theorem}
Now we prove this twisted duality using results of \cite{XY1},  \cite{XY2} and \cite{KMSW2}.
Fermion systems and quantum spin chains are formally equivalent via the Jordan-Wigner
 transformation.However this is not mathematically precise as the Jordan-Wigner
transformation contains an infinite product of Pauli spin matrices which may not converge
in the GNS spaces. We follow the idea of \cite{XY1}. 
First we introduce an automorphism $\Theta_{-}$ of $\frak A_{CAR}$ by the following equations: 
$$ \Theta_{-}(c^{*}_{j})= - c^{*}_{j} ,\:  \Theta_{-}(c_{j})= - c_{j}  \: ( j\leq 0), $$
$$ \Theta_{-}(c^{*}_{k})=  c^{*}_{k} ,\:  \Theta_{-}(c_{k})= c_{k} \: ( k >0 ) .$$
Let $\tilde{\frak A}$ be the crossed product of $\frak A_{CAR}$ by  the $\bfZ_{2}$ action $\Theta_{-}$ . 
$\tilde{\frak A}$ is the $C^{*}$-algebra generated by
$\frak A_{CAR}$ and a unitary $T$ satisfying 
$$ T=T^{*}, \: T^{2}= 1 , \: \: T Q T = \Theta_{-}(Q)\:\:\: ( Q \in \frak A_{CAR}).$$
Via the following formulae, we regard $\frak A$ as a subalgebra of $\tilde{\frak A}$:
\begin{align}
\sigma_z^{(j)}&=2c_j^*c_j -1
\nonumber\\
\sigma_x^{(j)}&=TS_{j}(c_j + c_j^*)
\nonumber\\
\sigma_y^{(j)}&=iTS_{j} (c_j - c_j^*).
\label{eqn:z6}
\end{align}
where
\begin{align*} 
S_{n}
=\left\{\begin{gathered}
\sigma_z^{(1)}\cdots\sigma_z^{(n-1)}\quad n>1\\
1\quad\quad\quad\quad n=1\\
\sigma_z^{(0)}\cdots\sigma_z^{(n)}\quad n<1.
\end{gathered}
\right.
\end{align*}
We extend the automorphism $\Theta$ of $\frak A_{CAR}$ to $\tilde{\frak A}$
via the  following equations:
$$ \Theta (T) = T  , \: \Theta ( \sigma_x^{(j)}) = - \sigma_x^{(j)} ,  \:
\Theta ( \sigma_y^{(j)}) = - \sigma_y^{(j)} , \: \Theta ( \sigma_z^{(j)}) =  \sigma_z^{(j)} .$$
As is the case of the CAR algebra, we set
$$(\frak A )_{\pm} = \{ Q \in \frak A |  \Theta (Q) = \pm Q \}, \:  
(\frak A_{\Lambda})_{\pm} = (\frak A)_{\pm} \cap \frak A_{\Lambda} , \:
(\frak A_{loc})_{\pm} = (\frak A)_{\pm} \cap \frak A_{loc} .$$ 
Then, it is easy to see that 
$$(\frak A)_{+} = (\frak A^{CAR})_{+} , \:  
(\frak A_{\Lambda})_{+} = (\frak A^{CAR}_{\Lambda})_{+} , \:
(\frak A_{loc})_{+} = (\frak A^{CAR}_{loc})_{+}  .$$ 
Let $\psi$ be a pure state of $\frak A_{CAR}$ and assume that $\psi$ is $\Theta$ invariant.
Let $\psi_{+}$ be the restriction of $\psi$ to  $ (\frak A^{CAR})_{+}=(\frak A)_{+} $.
$\psi_{+}$ is extendible to a $\Theta$ invariant state $\varphi_{0}$ of  $\frak A$
via the following formula:
\begin{equation}
 \varphi_{0} (Q)  =  \psi_{+}(Q_{+}), \quad  Q_{\pm} = \frac{1}{2} (Q\pm\Theta (Q)) \in (\frak A)_{\pm} .
\label{eqn:z7}
\end{equation}
In general,  $\varphi_{0}$  may not be a pure state but  if $\varphi$ is a pure state extension of  
$\psi_{+}$ to $\frak A$, the relation between $ \varphi_{0}$ and $\varphi$ is written as
$ \varphi_{0}(Q) =  \varphi (Q_{+}) $.
That $ \varphi_{0}$ and $\varphi$ are identical or not depends on existence of a unitary 
implementing $\Theta_{-}$ on $\frak H_{\psi}$. 
\begin{pro}
\label{pro:ext1}
Let $\psi$ be a $\Theta$ invariant pure state of $\frak A^{CAR}$ and $\psi_{+}$ be the restriction
of $\psi$ to  $(\frak A^{CAR})_{+}$.
\begin{description}
\item[(i)]
 Suppose that  $\psi$ and $\psi\circ\Theta_{-}$ are not unitarily equivalent.
The unique $\Theta$ invariant extension $\varphi$ of $\psi_{+}$ to $\frak A$ is a pure state. 
If $\psi$ is translationally invariant, $\varphi$ is translationally invariant as well.
\item[(ii)] 
 Suppose that  $\psi$ and $\psi\circ\Theta_{-}$ are unitarily equivalent and
that  $\psi_{+}$ and $\psi_{+}\circ\Theta_{-}$ are unitarily equivalent as states of
$(\frak A^{CAR})_{+}$.
The unique $\Theta$ invariant extension $\varphi$ of $\psi_{+}$ to $\frak A$ is a pure state. 
If $\psi$ is translationally invariant, $\varphi$ is translationally invariant as well.
\item[(iii)] 
 Suppose that  $\psi$ and $\psi\circ\Theta_{-}$ are unitarily equivalent and
that  $\psi_{+}$ and $\psi_{+}\circ\Theta_{-}$ are not unitarily equivalent as states of
$(\frak A^{CAR})_{+}$.
There exists  a pure state extension $\varphi$ of $\psi_{+}$ to $\frak A$ which is not
 $\Theta$ invariant. Furthermore,
 we can identify the GNS Hilbert spaces $\frak H_{\psi_{+}}$ and $\frak H_{\varphi}$ and
 \begin{equation}
 \pi_{\varphi}(\frak A)^{\dprime} =  \pi_{\varphi}((\frak A)_{+})^{\dprime}.
\label{eqn:z8}
 \end{equation}
If $\psi$ is translationally invariant, $\varphi$ is a periodic state with period 2,
$\varphi\circ\tau_{2} = \varphi$ and 
 \begin{equation}
 \pi_{\varphi}(\frak A_{L})^{\dprime} =  \pi_{\varphi}((\frak A_{L})_{+})^{\dprime}, \quad
  \pi_{\varphi}(\frak A_{R})^{\dprime} =  \pi_{\varphi}((\frak A_{R})_{+})^{\dprime}
 \label{eqn:z9}
 \end{equation}
where we set $(\frak A_{L,R})_{+} = (\frak A_{L,R}) \cap (\frak A)_{+}$.
\end{description}
\label{pro:duality21}
\end{pro}
\bigskip
\begin{pro}
\begin{description}
\item[(i)]
Let $\psi$ be a $\Theta$ invariant pure state of $\frak A^{CAR}$ and $\Lambda$ be a subset of
$\bfZ$. Then, the twisted Haag duality (\ref{eqn:z4}) holds for $\Lambda$ if and only if
\begin{equation}
\pi_{\psi_{+}}((\frak A^{CAR}_{\Lambda})_{+} )^{\dprime} = 
\pi_{\psi_{+}} ((\frak A^{CAR}_{\Lambda^{c}})_{+} )^{\prime} 
\label{eqn:z10}
\end{equation}
on the GNS space $\frak H_{\psi_{+}}$ associated with the state $\psi_{+}$ of  $(\frak A^{CAR})_{+}$ 
\item[(ii)]
Let $\varphi$ be a $\Theta$ invariant pure state of $\frak A$ and $\Lambda$ be a subset of
$\bfZ$. Then, the Haag duality holds for $\Lambda$ if and only if
\begin{equation}
\pi_{\varphi_{+}}((\frak A_{\Lambda})_{+} )^{\dprime} = 
\pi_{\varphi_{+}} ((\frak A_{\Lambda^{c}})_{+} )^{\prime} 
\label{eqn:z11}
\end{equation}
on the GNS space $\frak H_{\varphi_{+}}$ associated with the restriction $\varphi_{+}$ of   $\varphi$
to $(\frak A)_{+}$ .
\end{description}
\label{pro:duality22}
\end{pro}
Theorem \ref{th:dualityFermi} follows from the above Proposition \ref{pro:duality21} ,
Proposition \ref{pro:duality22} and the Haag duality for spin systems. 
\\
\\
{\it Proof of Proposition \ref{pro:duality21}}
\\
Set $X_{j} = c_{j} + c_{j}^{*}$. 
As $\psi$ is $\Theta$ invariant, the GNS space $\frak H_{\psi}$ is a direct sum of
$\frak H_{(\psi )}^{( \pm)}$ where
$$\frak H_{\psi}^{(+)} = \overline{ \pi_{\psi}((\frak A)_{+})\Omega} ,\quad         
\frak H_{\psi}^{(-)} = \overline{ \pi_{\psi}((\frak A)_{+} X_{j})\Omega}  .$$
The representation $\pi_{\psi}((\frak A)_{+})$  of $(\frak A)_{+}$ on $\frak H_{\psi}$  is 
decomposed into mutually disjoint irreducible representations on $\frak H_{\psi }^{(\pm)}$.
\par
Let $\psi$ and $\tilde{\psi}$ be $\Theta$ invariant states of ${\frak A}^{CAR}$. 
The argument in 2.8 of \cite{Voiculescu} shows that if $\psi_{+}$ and $\tilde{\psi}_{+}$ of
$(\frak A)_{+}$ are equivalent, $\psi$ and $\tilde{\psi}$  are equivalent.
Now we show (i). If pure states $\psi$ and  $\psi\circ\Theta_{-}$ are not equivalent,
 $\psi_{+}=\varphi_{+}$ is not equivalent to $(\varphi\circ\Theta_{-})_{+}$ and 
 $(\varphi\circ\Theta_{-}\circ Ad(X_{j}))_{+}$.
 Consider the GNS representation $\{\pi_{\varphi}(\frak A),\Omega_{\varphi}, \frak H_{\varphi}\}$
 of $\frak A$. If we restrict $\pi_{\varphi}$ to $(\frak A)_{+}$ it is the direct sum of two
 irreducible GNS representations associated with $\psi_{+}=\varphi_{+}$ and $(\varphi\circ\Theta_{-}\circ Ad(X_{j}))_{+}$.
 So we set 
 $$\frak H= \frak H_{\varphi} , \: \frak H= \frak H_{1}\oplus \frak H_{2}, \:
 \frak H_{1}= \frak H_{\varphi_{+}}, \:
  \frak H_{2}= \frak H_{(\varphi\circ\Theta_{-}\circ Ad(X_{j}))_{+}} .$$
 Any bounded operator $A$ on $\frak H$ is written in a matrix form,
 \begin{equation}
 A = \left(\begin{array}{cc}a_{11} & a_{12} \\a_{21} & a_{22}\end{array}\right)
 \label{eqn:z12}
 \end{equation}
 where $a_{11} $(resp. $a_{22} $) is a bounded operator on $\frak H_{1}$ (resp. $\frak H_{2}$)
 and $a_{12} $(resp. $a_{21} $) is a bounded operator from $\frak H_{2}$ to $\frak H_{1}$
 (resp. a bounded operator from $\frak H_{1}$ to $\frak H_{2}$.
 As $\psi_{+}=\varphi_{+}$ is not equivalent to $(\varphi\circ\Theta_{-}\circ Ad(X_{j}))_{+}$,
  \begin{equation}
 P = \left(\begin{array}{cc} a & 0 \\ 0 &b \end{array}\right) 
 \label{eqn:z13}
 \end{equation} 
is an element of $\pi_{\varphi} ((\frak A)_{+} )^{\dprime}$ and 
$\pi_{\varphi}(\sigma_{x}^{(j)})$ looks like
\begin{equation}
\pi_{\varphi}(\sigma_{x}^{(j)} ) = \left(\begin{array}{cc} 0 & d \\ d^{*} &0 \end{array}\right) 
\label{eqn:z14}
 \end{equation}
A direct computation shows that an operator $A$ of the matrix form  (\ref{eqn:z12}) 
commuting with (\ref{eqn:z13}) and  (\ref{eqn:z14}) is trivial. This shows that the state $\varphi$ is pure.
\\
The translational invariance of $\varphi$ follows from  translational invariance of $\psi$
and $\varphi (Q) = \psi (Q_{+})$.
 \par
(ii) of Proposition \ref{pro:duality21} can be proved by constructing the representation
of $\frak A$ on the GNS space of Fermion.
By our assumption, $\pi_{\psi_{+}}((\frak A)_{+})$ is not equivalent to
$\pi_{\psi_{+}}(Ad(X_{j}) (\frak A)_{+})$. Hence $\pi_{\psi_{+}}((\frak A)_{+})$  is equivalent
to $\pi_{\psi_{+}}(\Theta_{-}(\frak A)_{+}))$ and $\pi_{\psi_{+}}(Ad(X_{j})(\frak A)_{+}))$  is equivalent
to $\pi_{\psi_{+}}(\Theta_{-}(Ad(X_{j})\frak A)_{+}))$. It turns out that 
there exists a selfadjoint unitary $U(\Theta_{-})$ 
($U(\Theta_{-})^{*}=U(\Theta_{-})$, $U(\Theta_{-})^{2}=1$)  
on $\frak H_{\psi}$ such that
\begin{equation}
U(\Theta_{-}) \pi_{\psi}(Q) U(\Theta_{-})^{*},  \quad
U(\Theta_{-}) \in \pi_{\psi}((\frak A)_{+})^{\dprime}
\label{eqn:z15}
 \end{equation}
for any $Q$ in $\frak A^{CAR}$.
Any element $R$ of $\frak A$ is writtten in terms of fermion operators and $T$ as follows:
\begin{equation}
R =  R_{+} + TR_{-}   , 
 \label{eqn:z16}
 \end{equation}
where
$$ R_{+} = \frac{1}{2} ( R + \Theta (R)) \in (\frak A^{CAR})_{+} , \quad
 R_{-} = \frac{1}{2} ( TR - T\Theta (R)) \in (\frak A^{CAR})_{-} .$$
Using this formula, for any $R$ in $\frak A$, we set
\begin{equation}
\pi (R) =   \pi_{\psi} (R_{+}) + U(\Theta_{-})   \pi_{\psi} (R_{-} )
 \label{eqn:z17}
 \end{equation}
$\pi (R)$ gives rise to a representation of $\frak A$ on $\frak H_{\psi}$ and we set
\begin{equation}
\varphi (R) =  \left( \Omega_{\psi},   \pi (R) \Omega_{\psi}\right)  .
 \label{eqn:z18}
 \end{equation}
The representation $\pi (\frak A)$ is irreducible because
$\pi ((\frak A)_{+})^{\dprime}$ contains  $U(\Theta_{-})$ and hence $\pi (\frak A)^{\dprime}$
contains $\pi ((\frak A^{CAR})_{-})$ and  $\pi (\frak A)^{\dprime}= \frak B (\frak H_{\varphi})$.
\\
As in (i), the translational invariance of $\varphi$ follows  from $\Theta$ invariance of $\varphi$
(by construction) and translational invariance of $\psi$ .
\par
To show (iii), we construct an irreducible representation of $\frak A$ on 
the GNS space $\frak H_{+} =\overline{ \pi_{\psi_{+}}((\frak A^{CAR})_{+} )\Omega_{\psi}}$.
Now under our assumption there exists a selfadjoint unitary $V(\Theta_{-})$ satisfying
\begin{equation}
V(\Theta_{-}) \pi_{\psi}(Q) V(\Theta_{-})^{*} = \pi_{\psi}(\Theta_(Q)),  \quad
 V(\Theta_{-}) \in \overline{\pi_{\psi}((\frak A)_{-})}^{w}
\label{eqn:z19}
 \end{equation}
for any $Q$ in $\frak A^{CAR}$. For $R$ written in the form (\ref{eqn:z16}), we set 
\begin{equation}
\pi (R) =   \pi_{\psi} (R_{+}) + V(\Theta_{-})   \pi_{\psi} (R_{-} )
 \label{eqn:z20}
 \end{equation}
for $R$ in $\frak A$ and $\pi (R)$ belongs to the even part $\pi_{\psi}((\frak A^{CAR})_{+})^{\dprime}$.
and $\pi (\frak A)$ acts irreducibly on $\frak H_{+}$.
\\
To show periodicity of the state $\varphi$, we introduce a unitary $W$
satisfying
$$W\Omega_{\psi}=\Omega_{\psi} , \quad 
W \pi_{\psi} (Q) W^{*} =  \pi_{\psi} (\tau_{1}(Q))  , \quad Q \in \frak A^{CAR}$$
The adjoint action of both unitaries $W V(\Theta_{-}) W^{*}$ and 
$V(\Theta_{-}) \pi_{\psi}(\sigma_{z}^{(1)})$ gives rise to the same automorphism on 
$\pi_{\psi}(\frak A^{CAR})$. 
By irreducibility of the representation $\pi_{\psi}(\frak A^{CAR})$, 
$W V(\Theta_{-}) W^{*}$ and $V(\Theta_{-}) \pi_{\psi}(\sigma_{z}^{(1)})$ 
differ in a phase factor.
\begin{equation}
W V(\Theta_{-}) W^{*} = c V(\Theta_{-}) \pi_{\psi}(\sigma_{z}^{(1)}) 
 \label{eqn:z21}
 \end{equation}
where $c$ is a complex number with $|c| =1$ .
 As both sides in (\ref{eqn:z21}) are selfadjoint , $c = \pm 1$.
Then, 
\begin{equation}
W^{2} V(\Theta_{-}) (W^{2})^{*} =  V(\Theta_{-}) \pi_{\psi}(\sigma_{z}^{(1)}\sigma_{z}^{(2)})
 \label{eqn:z22}
 \end{equation}
 This implies that the state $\varphi$ is periodic, for example,
 \begin{eqnarray*}
 \varphi (\tau_{2}(\sigma_{x}^{(1)}))  &=& 
 \left( \Omega_{\psi}, ( W^{2} V(\Theta_{-}) \pi_{\psi}(c_{1}+c_{1}^{*}) (W^{2})^{*}  \Omega_{\psi}\right)
 \\
 &=&\left( \Omega_{\psi},  V(\Theta_{-})  \pi_{\psi}((\sigma_{z}^{(1)}\sigma_{z}^{(2)})(c_{3}+c_{3}^{*}))  \Omega_{\psi}\right)
 \\
&=& \varphi (\tau_{2}(\sigma_{x}^{(3)})) .
 \end{eqnarray*}
 {\it End of Proof of Proposition \ref{pro:duality21}}
\begin{rmk}
In \cite{XXZ}, using expansion technique(but not the exact solution) 
we have shown  the XXZ Hamiltonian $H_{XXZ}$ with large Ising type anisotorpy 
$\Delta >>1$ 
$$H_{XXZ} = \sum_{j=-\infty}^{\infty} \{ \Delta \sigma_{z}^{(j)}\sigma_{z}^{(j+1)} +
\sigma_{x}^{(j)}\sigma_{x}^{(j+1)} + \sigma_{y}^{(j)}\sigma_{y}^{(j+1)} \}$$
has exactly two pure ground states 
$\varphi$ and 
$$\varphi\circ\Theta=\varphi\circ\tau_{1} \ne \varphi .$$ 
The unique $\Theta$ invariant ground state  $(1/2\varphi +\varphi\circ\tau_1)$ is 
a pure state of $(\frak A)_{+}$
In this example, the phase factor $c$ of (\ref{eqn:z21}) is $-1$. 
\end{rmk}
\bigskip
\noindent
\newline
{\it Proof of Proposition \ref{pro:duality22}}
\\
We now prove (i). Suppose that $\psi$ is a $\Theta$ invariant pure state of $\frak A$. 
\\
Let  $\{\pi_{\psi}(\frak A^{CAR}), \Omega_{\psi}, \frak H_{\psi}\}$ 
be the GNS triple associated with $\psi$ and $U$ be the selfadjoint unitary satisfying
$$U\pi_{\psi}(Q)U^{*} = \pi_{\psi}(\Theta (Q)) ,  \quad U\Omega_{\psi} = \Omega_{\psi} .$$
We set
$$  \frak H_{\pm} =\{ \xi \in \frak H_{\psi} \: |   U\xi = \pm \xi \} $$
and let $P_{\pm}$ be the projection to $ \frak H_{\pm}$.
\par
First we assume (\ref{eqn:z10}) and fix $k$ in $\Lambda$ and $l$ in $\Lambda^{c}$. 
Any element  $Q$ in the commutant of $\pi_{\psi} (\frak A^{CAR})$ is written as
\begin{equation} 
Q = Q_{1} + Q_{2} Z_{l} , \quad  Z_{l} =U \pi_{\psi}((c_{l} + c^{*}_{l}))
\label{eqn:z23}
\end{equation}
where 
$$Q_{1} = \frac{1}{2}(  Q + UQU^{*} ) , \quad Q_{2} = \frac{1}{2}(  Q - UQU^{*} ) Z_{l}^{*} .$$
It is easy to see that $UQ_{1}U^{*} = Q_{1}$, $U Q_{2}U^{*}= Q_{2}$, and that
 $Q_{1}, Q_{2}$ is in $\pi_{\psi} (\frak A^{CAR}_{\Lambda})^{\prime}$.
 It turns out that, to prove our claim, it suffices to show that an operator $Q$ commuting with $U$ and 
 $ \pi_{\psi} (\frak A^{CAR}_{\Lambda})$ is in the weak closure of 
 $\pi_{\psi}( (\frak A^{CAR}_{\Lambda^{c}})_{+})$.
 \par
 Now let $Q$ be an operator satisfying
 $ [Q, U] = 0 , \quad  [Q,  \pi_{\psi} (R) ] =0 $
 for any  $R$  in $\pi_{\psi} (\frak A^{CAR}_{\Lambda})$.
Set 
$$ Q_{\pm} =  P_{\pm} Q P_{\pm} .$$ 
Due to our assumption (\ref{eqn:z10}), we obtain
\begin{equation}
Q_{+}   = w-\lim_{\alpha}   P_{+}\pi_{\psi} ( Q_{\alpha})P_{+}
\label{eqn:z24}
\end{equation}
for a sequence $Q_{\alpha} $ in $(\frak A^{CAR}_{\Lambda^{c}})_{+}$.
As $Q$ commutes with the selfadjoint unitary 
$\overline{X}_{k} = \pi_{\psi}( c_{k} + c^{*}_{k})$,
we get
\begin{equation}
Q_{-} =    P_{-} \overline{X}_{k}  Q_{+} \overline{X}_{k} P_{-} 
=P_{-} \overline{X}_{k} P_{+} Q P_{+} \overline{X}_{k} P_{-} 
=P_{-} \overline{X}_{k}  Q \overline{X}_{k} P_{-} .
\label{eqn:z25}
\end{equation}
Inserting (\ref{eqn:z24}) in (\ref{eqn:z25}) we arrive at
\begin{eqnarray}
Q_{-} &=& w-\lim_{\alpha}  P_{-}  \overline{X}_{k} \pi_{\psi} ( Q_{\alpha})   \overline{X}_{k} P_{-}
\nonumber\\
&=& w-\lim_{\alpha}  P_{-}  \pi_{\psi} (( c_{k} + c^{*}_{k}) Q_{\alpha}( c_{k} + c^{*}_{k}))  P_{-}
\nonumber\\
&=& w-\lim_{\alpha}  P_{-}  \pi_{\psi} (Q_{\alpha})  P_{-}
\label{eqn:z26}
\end{eqnarray}
where we used the conditions that $( c_{k} + c^{*}_{k})\in (\frak A^{CAR}_{\Lambda})$ and 
that $Q_{\alpha}^{+} \in (\frak A^{CAR}_{\Lambda^{c}})_{+}$.
(\ref{eqn:z24}) and (\ref{eqn:z25}) imply that
\begin{equation}
Q  = w-\lim_{\alpha} \pi_{\psi} ( Q_{\alpha}) \in 
\pi_{\psi}( \frak A^{CAR}_{\Lambda^{c}})_{+})^{\dprime}
\label{eqn:z27}
\end{equation}
(\ref{eqn:z27}) is the property we claimed.
\\
\\
Next we show (\ref{eqn:z11}) assuming twisted Haag duality (\ref{eqn:z11}) . We use the same notation
as above. 
\par
The representation $ \pi_{\psi}$ restricted to $(\frak A^{CAR}_{\Lambda})$
is a direct sum of representations $\pi^{\pm} $ where
$$\pi^{\pm} ((\frak A^{CAR}_{\Lambda})_{+}) ) 
=  P_{\pm} \pi_{\psi}( (\frak A^{CAR}_{\Lambda})_{+})) P_{\pm}$$ 
on $\frak H_{\pm}$. 
We denote $\tilde{\pi}^{\pm} $ by the representation of $(\frak A^{CAR}_{\Lambda^{c}})_{+}$ 
on $\frak H_{\pm}$.
$\pi^{\pm} $ of $(\frak A^{CAR}_{\Lambda})_{+}$ 
are mutually unitarily equivalent because the operator $Z_{l}$ interwtines these representations.
The same is true for $\tilde{\pi}^{\pm} $ for $(\frak A^{CAR}_{\Lambda^{c}})_{+}$.
Let $\frak M_{\pm}$ be the von Neumann algebra on $\frak H_{\pm}$ generated by
$\pi^{\pm} ((\frak A^{CAR}_{\Lambda})_{+})$ 
As $\pi^{\pm}$ are unitarily equivalent, $\Xi = Ad(\pi_{\psi} ((c_{k}+c^{*}_{k})))$ gives rise to 
an automorphism of  $\frak M_{\pm}$.  Thus   $ Ad(\pi_{\psi} ((c_{k}+c^{*}_{k}))$ is
an automorphim of the commutant ${\frak M_{\pm}}^{\prime}$ on $\frak H_{\pm}$.
\par 
Now suppose $Q_{+}$ is an element of ${\frak M_{+}}^{\prime}$ on $\frak H_{+}$
and we have to show that $Q_{+}$ is in 
$\tilde{\pi}^{+} ((\frak A^{CAR}_{\Lambda^{c}})_{+})^{\dprime}$.
\par
Set $X_{k} = (c_{k}+c^{*}_{k}$ and
\begin{equation}
Q = P_{+}Q_{+}P_{+} + P_{-} \pi_{\psi} (X_{k})Q_{+} \pi_{\psi} (X_{k})P_{-} .
\label{eqn:z28}
\end{equation}
Then, we claim that $Q$ commutes with $(\pi_{\psi}(\frak A^{CAR}_{\Lambda})$.
To see this, first take $R$ from   $(\frak A^{CAR}_{\Lambda})_{+}$ and we obtain
\begin{eqnarray}
 &&Q \pi_{\psi}(R) =
 P_{+}Q_{+} \pi^{+}(R) P_{+} +  P_{-} \pi_{\psi} (X_{k})  Q_{+} 
\pi_{\psi} (X_{k} R X_{k}) \pi_{\psi}(X_{k})P_{-}
\nonumber\\
&&
= P_{+}\pi^{+}(R)Q_{+}  P_{+} +  P_{-} \pi_{\psi} (X_{k}) P_{+} Q_{+} P_{+}\pi_{\psi} (X_{k} R X_{k})P_{+} \pi_{\psi}(X_{k})P_{-}
\nonumber\\ 
 &&= P_{+}\pi^{+}(R)Q_{+}  P_{+} +  P_{-}(\pi_{\psi} (X_{k}) P_{+} \pi_{\psi} (X_{k} R X_{k}) P_{+}Q_{+} P_{+} \pi_{\psi}(X_{k})P_{-}
\nonumber\\
&&= \pi_{\psi}(R) Q .
\label{eqn:z29}
\end{eqnarray}
On the other hand,
\begin{eqnarray}
&& \pi_{\psi}(X_{k})Q \pi_{\psi}(X_{k})=  P_{+} \pi_{\psi}(X_{k})Q \pi_{\psi}(X_{k}) P_{+}
+ P_{-} \pi_{\psi}(X_{k})Q \pi_{\psi}(X_{k}) P_{-}
\nonumber\\
&&=
 P_{+} \pi_{\psi}(X_{k}) P_{-}Q P_{-} \pi_{\psi}(X_{k}) P_{+}
+ P_{-} \pi_{\psi}(X_{k}) P_{+} Q P_{+}  \pi_{\psi}(X_{k}) P_{-}
\nonumber\\
&&=
 P_{+} \pi_{\psi}(X_{k})\pi_{\psi}(X_{k}) Q_{+} \pi_{\psi}(X_{k})\pi_{\psi}(X_{k}) P_{+}
+ P_{-} \pi_{\psi}(X_{k}) Q_{+}  \pi_{\psi}(X_{k}) P_{-}
\nonumber\\
&&=
P_{+}Q_{+})P_{+} + P_{-} \pi_{\psi}(X_{k}) Q_{+}  \pi_{\psi}(X_{k}) P_{-}
\nonumber\\
&&= Q.
\label{eqn:z30}
\end{eqnarray}
As a consequence, 
$$Q \in \pi_{\psi}(\frak A^{CAR}_{\Lambda})^{\prime} 
= \pi_{\psi}(\frak A^{CAR}_{\Lambda^{c}})^{\dprime} ,\quad
Q_{+} \in \pi^{+}((\frak A^{CAR}_{\Lambda^{c}})_{+})^{\dprime}$$ 
\\
\\
As (ii) can be shown in the same manner, we omit the detail.
\\ 
{\it End of Proof of Proposition \ref{pro:duality22}}

\section{Split Property and Spectral Gap}\label{SecSplit}
\setcounter{theorem}{0}
\setcounter{equation}{0}
Once Haag duality is proven, it is possible to show that the presence of the spectral gap
implies split property in the sense of S.Doplicher and R.Longo. (cf.\cite{DoplicherLongo}) 
This result is known in case of the relativistic QFT case. We explain the proof rather briefly. 
In our proof we use results on maximal violation of Bell's inequality due to
Stephen J.Summers and Reinhard Werner in \cite{SummersWerner} .
\\
\\
First let us recall the definition of split property or split inclusion.
Let $\frak M_{1}$ and $\frak M_{2}$ be a commuting pair of factors acting on a Hilbert space $\frak H$,
 $\frak M_{1} \subset \frak M_{2}^{\prime}$. We say the inclusion is split
if  there exists an intermediate type I factor $\calN$ such that 
\begin{equation}
\frak M_{1} \subset \calN \subset \frak M_{2}^{\prime} \subset \frak B (\frak H )
\label{eqn:v1}
\end{equation}
The split inclusion is used for analysis of local QFT and of von Neumann algebras and
 some general feature of this concept is investigated for abstract von Neumann alegebras.
by J.von Neumann and later by S.Doplicher and R.Longo in \cite{DoplicherLongo} .
R.Longo used this notion of splitting for his solution to the factorial 
Stone-Weierstrass conjecture in \cite{Longo1}. 
\par
If (\ref{eqn:v1}) is valid, the inclusion of the type I factors 
$\calN =  \frak B (\frak H_{1}) \subset \frak B (\frak H )$ implies factorization of the underlying Hilbert 
spaces and we obtain $\frak H = \frak H_{1} \otimes \frak H_{2}$ and tensor product
\begin{equation}
\frak M_{1}= \tilde{\frak M}_{1}\otimes  1_{\frak H_{2}} 
\subset \frak B (\frak H_{1})\otimes 1_{\frak H_{2}} , \:\:
\frak M_{2}= 1_{\frak H_{1}} \otimes \tilde{\frak M}_{2}  
\subset 1_{\frak H_{1}} \otimes \frak B (\frak H_{2}).
\label{eqn:v2}
\end{equation}
In this sense the split inclusion is statistical independence of two algebras
$\frak M_{1}$ and $\frak M_{2}$.
\par
When $\frak M_{2}$ is the commutant  $\frak M_{1}^{\prime}$.of $\frak M_{1}$,
the split property of the inclusion    $\frak M_{1} \subset \frak M_{2}^{\prime}$
is nothing but  the condition that $\frak M_{1}$ and hence  $\frak M_{2}$  are type I von Neumann
algebras .
In our case of quantum spin chains,  we set 
$\frak M_{1} = \frak M_{R}=\pi_{\varphi} ( \frak A_{R})^{\dprime} $, and 
$\frak M_{2} = \frak M_{L}=\pi_{\varphi} (\frak  A_{L})^{\dprime} $.
When the state $\varphi$ is translationally invariant and pure, $\frak M_{2}$ is the commutant 
of $\frak M_{1}$ due to Haag dualtiy.
\\
\\ 
In 1987, Stephen J.Summers and Reinhard Werner found the characterization of
split property in terms of violation of Bell's inequality.
We now explain their results  in \cite{SummersWerner} .
Fix  a commuting pair of factors $\frak M_{1}$ and $\frak M_{2}$ and 
let $\frak M$ be the von Neumann algebra generated by  $\frak M_{1}$ and $\frak M_{2}$,
$\frak M = \frak M_{1}\vee  \frak M_{2}$ and let$\varphi$ be a normal state of $\frak M$ .
\par
By an admissible quadraple $I = \{ X_{1}, X_{2}, Y_{1}, Y_{2}\}$, 
we mean a quartet of operators $X_{1}$, $X_{2}$ in  $\frak M_{1}$ and  $Y_{1}$, $Y_{2}$ in  $\frak M_{2}$ satisfying 
$$-1 \leq X_{1}\leq 1 , \: -1 \leq X_{2}\leq 1, \quad -1 \leq Y_{1}\leq 1 , \: -1 \leq Y_{2}\leq 1.$$
We set
\begin{equation}
\beta (\varphi ,  \frak M_{1}, \frak M_{2}) = 
\frac{1}{2}\sup_{I} \varphi (X_{1}(Y_{1}+Y_{2})+X_{2}(Y_{1}-Y_{2}))
 \label{eqn:v3}
\end{equation}
where the supremun is taken in all admissible quadraple $I = \{ X_{1}, X_{2}, Y_{1}, Y_{2}\}$ . 
We call $\beta (\varphi ,  \frak M_{1}, \frak M_{2}) $ {\it the Bell's constant}.
\par
The following results are known. (cf. \cite{SummersWerner2}):
\begin{description}
\item[(i)] 
$1\leq \beta (\varphi ,  \frak M_{1}, \frak M_{2}) \leq \sqrt{2}$
\item[(ii)]
If either $\frak M_{1}$ or,$\frak M_{2}$ is commutative, $\beta (\varphi ,  \frak M_{1}, \frak M_{2})=1$.
\item[(iii)]
If the normal state $\varphi$ of $\frak M$ is a convex combination of product states, then
$\varphi = \sum_{i}  \psi_{1}^{(i)}\otimes \psi_{2}^{(i)}$, $\beta (\varphi ,  \frak M_{1}, \frak M_{2})=1$.
\item[(iii)]
If $X_{1}$ and $X_{2}$ attain the maximum value $\sqrt{2}$ of the Bell's constant ,
$\beta (\varphi ,  \frak M_{1}, \frak M_{2})=\sqrt{2}$, then,
\begin{equation}
\varphi (X_{i}^{2}Q) = \varphi (QX_{i}^{2})=\varphi (Q), \quad \varphi ((X_{1}X_{2}+X_{2}X_{1})Q)=0
\label{eqn:v4}
\end{equation}
for $i=1,2$ and for any $Q$ in $\frak M_{1}$.
\end{description}
When the state $\varphi$ is faithful on  $\frak M_{1}$, the equation (\ref{eqn:v4}) means that 
$\sigma_{x}=X_{1}$, $\sigma_{y}=X_{2}$, and $\sigma_{z}= iX_{1}X_{2}$ satisfy
the relation of Pauli matrices and that the state $\varphi$ restricted to 
these Pauli spin matrices is the tracial state. 
If the maximum value $\sqrt{2}$ of the Bell's constant is not attained by some elements,
 it is possible to find a sequence of operators asymptotically satisfying the relation 
 of Pauli matrices in the ultra product of $\frak M_{1}$. Then, by applying a result of  strong stability
 of von Neumann algebras due to A.Connes, 
we are led to the followin relation between split property, 
strong sability of von Neumann algebras and Bell's constant.
(See \cite{SummersWerner} for proof.)
\begin{theorem}[S. J.Summers and Reinhard Werner]
\label{th:SummersWerner}
Let $\frak M$ be a von Neumann algebra in a separable Hilbert space $\frak H$ with  
cyclic separating vector. 
The following conditions are equivalent.
\\
(i)  $\frak M$ is strongly stable, i.e.
$$\frak M \cong \frak M \otimes \frak R_{1}$$
 where $\frak R_{1}$ is the hyperfinite $II_{1}$ factor. 
\\
(ii) For every normal state $\varphi$ of $\frak B(\frak H)$, $ \beta (\varphi ,  \frak M, \frak M^{\prime}) = \sqrt{2} .$
\end{theorem}
\begin{cor}
\label{cor:GapSplit}
(i) Let $\varphi$ be a translationally invariant pure state of $\frak A$ and set
\begin{equation} 
C_{j} = \sup  | \varphi (Q \tau_{j}(R)) -\varphi(Q)\varphi(R) | 
\label{eqn:v5}
\end{equation}
where the supremum is taken for $Q \in  \frak A_{L}$, and $R  \in  \frak A_{R}$
satisfying 
\\
$\norm{Q}\leq 1 ,  \norm{R}\leq 1$  .
\par
Suppose that the following uniform decay of correlation is valid.
\begin{equation}
\lim_{j \to \infty} C_{j} = 0
\label{eqn:v6}
\end{equation}
Then, $\frak M_{L}$ and $\frak M_{R}$  are of type I.
\\
(ii) Let $\psi$ be a translationally invariant pure state of $\frak A^{CAR}$ and set
\begin{equation} 
C_{j} = \sup  | \varphi (Q \tau_{j}(R)) -\varphi(Q)\varphi(R) | 
\label{eqn:v7}
\end{equation}
where the supremum is taken for $Q \in  \frak A^{CAR}_{L}$, and $R  \in  \frak A^{CAR}_{R}$
satisfying 
\\
$\norm{Q}\leq 1 ,  \norm{R}\leq 1$  .
\par
Suppose that the following uniform decay of correlation is valid.
\begin{equation}
\lim_{j \to \infty} C_{j} = 0
\label{eqn:v8}
\end{equation}
Then, $\frak M^{CAR}_{L} =\pi_{\psi}(\frak A^{CAR}_{L})^{\dprime}$ and 
$\frak M^{CAR}_{R} =\pi_{\psi}(\frak A^{CAR}_{R})^{\dprime}$  are of type I.
\end{cor}
{\it Proof of Corollary  \ref{cor:GapSplit}.}
To show the above corollary  \ref{cor:GapSplit} (i), first take $j$ large such that
$C_{j} < \epsilon $ and we have
\begin{equation}
 | \left( \Omega_{\varphi}, Q R\Omega_{\varphi}\right) -
\left( \Omega_{\varphi},Q \Omega_{\varphi}\right)  \left( \Omega_{\varphi}, R\Omega_{\varphi}\right)  |
 < \epsilon || Q || \cdot || R||
 \label{eqn:v9}
\end{equation}
for any $Q$ in $\frak M_{(-\infty,0]}$ and any $R$ in $\frak M_{[j, \infty)}$. 
Let $\tilde{\varphi}$ be the vector state associated with $\Omega_{\varphi}$ and
restrict it to  $\frak M_{(-\infty,0] \cup [j, \infty)}$.
Then, $\tilde{\varphi}_{(-\infty,0] \cup [j, \infty)}$ is close to a product state
due to (\ref{eqn:v9})
\begin{equation}
\beta ( \tilde{\varphi}_{(-\infty,0] \cup [j, \infty)} , \frak M_{[j, \infty)}, \frak M_{(-\infty,0]}) \leq 1+ 2\epsilon 
 \label{eqn:v10}
\end{equation}
As the state $\varphi$ is pure, the von Neumann algebra $\frak M_{(-\infty,0] \cup [j, \infty)}$
 is type $I$ and by Haag duality explained in the previous section, we have
 $$ \frak M_{(-\infty,0] \cup [j, \infty)} \cap \frak M_{(-\infty,0]}^{\prime} =   \frak M_{[j, \infty)} .$$ 
The state $\tilde{\varphi}_{(-\infty,0] \cup [j, \infty)}$ may not be faithful. We reduce
$\frak M_{[j, \infty)}$ by the support projection $P$ for $\tilde{\varphi}_{(-\infty,0] \cup [j, \infty)}$. 
 We set 
$\frak M = P\frak M_{[j, \infty)}P$ and we apply Theorem \ref{th:SummersWerner} 
of S.J.Summes and R.Werner. 
As a result,  $\frak M$ is not strong stable. By construction,  $\frak M$ is hyperfinite, 
so $\frak M$ and $\frak M_{[j, \infty)}$ are  type $I$ von Neumann algebra.
(See Section 4 and Appendix of \cite{KMSW}.)
As $\frak M_{1j, \infty)}$ is the tensor product of a matrix algbera and $\frak M_{[j, \infty)}$,
it is of type $I$ as well.

The case of the corollary  \ref{cor:GapSplit} (ii) can be handle in the same way. 
Then, instead of (\ref{eqn:v9}) , we obtain
\begin{equation}
 | \left( \Omega_{\psi}, Q R\Omega_{\psi}\right) -
\left( \Omega_{\psi},Q \Omega_{\psi}\right)  \left( \Omega_{\psi}, R\Omega_{\psi}\right)  |
 < \epsilon || Q || \cdot || R||
 \label{eqn:v11}
\end{equation}
for for any $Q$ in $\tilde{\pi}(\frak A_{(-\infty,0]}^{CAR})^{\dprime}$ and 
any $R$ in $\pi (\frak A_{[j, \infty)}^{CAR})^{\dprime}$. 

As before, we express any element $Q$ of $\frak A^{CAR}$  as a sum of even and odd elements. 
$$Q = Q_{+} + Q_{-} ,  \:\:  Q_{\pm} \in \frak A^{CAR})_{\pm}, \:\: || Q_{\pm} || \leq ||Q|| .$$ 
The state $\psi$ is $\Theta$ invariant, and we see
$$\psi (QR) - \psi (Q)\psi (R) =  \psi (Q_{+}R_{+}) - \psi (Q_{+})\psi (R_{+})  
+  \psi (Q_{-}R_{-}) .$$
For $Q$ in $\frak A_{[1, \infty)}^{CAR}$ and $R$ in $\frak A_{(-\infty , 0]}^{CAR}$ 
\begin{eqnarray}
&& | \left(\Omega_{\psi}, \tilde{\pi}_{\psi}(R) \pi_{\psi}(Q)\Omega_{\psi}\right)
-\left(\Omega_{\psi}, \tilde{\pi}_{\psi}(R) \Omega_{\psi}\right)
\left(\Omega_{\psi},  \pi_{\psi}(Q)\Omega_{\psi}\right)|
\nonumber
\\
\leq &&
| \left(\Omega_{\psi}, \pi_{\psi}(R_{+}) \pi_{\psi}(Q_{+})\Omega_{\psi}\right)
-\left(\Omega_{\psi}, \pi_{\psi}(R_{+}) \Omega_{\psi}\right)
\left(\Omega_{\psi},  \pi_{\psi}(Q_{+})\Omega_{\psi}\right)|
\nonumber
\\
+&&| \left(\Omega_{\psi}, \pi_{\psi}(R_{-}) \pi_{\psi}(Q_{-})\Omega_{\psi}\right)
 \label{eqn:v12}
\end{eqnarray}
Thus we obtain the following estimate of the Bell's constant
\begin{equation}
\beta ( \tilde{\psi}_{(-\infty,0] \cup [j, \infty)} , \pi (\frak A_{[j, \infty)}^{CAR})^{\dprime},
\tilde{\pi}(\frak A_{(-\infty,0]}^{CAR})^{\dprime}) \leq 1+ 4\epsilon .
 \label{eqn:v13}
\end{equation}
(\ref{eqn:v13}) shows that  $\pi (\frak A_{[1, \infty)}^{CAR})^{\dprime}$ is of type $I$.
{\it End of Proof of Corollary  \ref{cor:GapSplit}.}
\bigskip
\bigskip
\noindent
\par
By setting $C_{j} = C_{0} e^{-M |j|}$ the above corollary \ref{cor:GapSplit} implies
Theorem \ref{th:mainSplit} (i).

We consider fermionic systems. A state $\psi$ of $\frak A^{CAR}$
or $\frak A_{\Lambda}^{CAR}$ is called even if $\psi\circ\Theta =\psi$.
Suppose that states $\psi_{1}$ of  $\frak A_{\Lambda}^{CAR}$
and  $\psi_{2}$ of  $\frak A_{\Lambda^{c}}^{CAR}$ are given and
that $\psi_{1}$ is even. We construct the graded tensor product state
$\psi_{1}\otimes_{\bfZ_{2}}\psi_{2}$ in the following manner.
Let $\{\pi_{k}(\cdot ) , \Omega_{k} , \frak H_{k}\}$ ($k=1,2$) be the GNS representation
associated with $\psi_{k}$.
As $\psi_{1}$ is even, there exists a selfadjoint unitary $\Gamma$ on $\frak H_{1}$
implementing $\Theta$ on $\frak A_{\Lambda}^{CAR}$: 
$$ \Gamma \pi_{1}(Q) \Gamma^{*} =\pi_{1}(\Theta (Q)) ,\quad   Q \in  \frak A_{\Lambda}^{CAR}$$
We introduce a representation $\pi $ of $\frak A^{CAR}$ on $\frak H = \frak H_{1}\otimes \frak H_{2}$
via the following identity:
$$  \pi (c_{j}) =  \pi_{1}(c_{j})\otimes 1 , \:\:  \pi (c_{j}) =  \pi_{2}  \Gamma \otimes (c_{k})$$
for $j$ in $\Lambda$ and $k$ in $\Lambda^{c}$.
We define $\psi_{1}\otimes_{\bfZ_{2}}\psi_{2}$ as the vector state for 
$\Omega_{1}\otimes\Omega_{2}$.
 $$\psi_{1}\otimes_{\bfZ_{2}}\psi_{2} (Q) = 
 \left( \Omega_{1}\otimes\Omega_{2} , \pi (Q) \Omega_{1}\otimes\Omega_{2}\right) .  $$
If $\psi$ is an even state of $\frak A^{CAR}$ and if the restriction of $\psi$ to 
$\frak A_{\Lambda}^{CAR}$ gives rise to a type $I$ representation,  $\psi$ is 
equivalent to $\psi_{1}\otimes_{\bfZ_{2}}\psi_{2}$ where $\psi_{1}$ is a even state
  of  $\frak A_{\Lambda}^{CAR}$ and  $\psi_{2}$ is a state of  $\frak A_{\Lambda^{c}}^{CAR}$. 
\par
Noticing these facts we see that the corollary \ref{cor:GapSplit} implies Theorem \ref{th:mainSplit} (ii).
\newpage
\section{$U(1)$ Gauge Symmetry}\label{SecCuntz}
\setcounter{theorem}{0}
\setcounter{equation}{0}
To complete our proof of Theorem \ref{th:main}, we use the main theorem of \cite{Split}.
and the proposition below.
\begin{theorem}
\label{th:Matsui2001}
Suppose that the spin $S$ of one site algebra $M_{2S+1}$($n=2S+1$) for $\frak A$ is
$1/2$.
Let $\varphi$ be a translationally invariant pure state of $\frak A$ such that
$\varphi_{R}$ gives rise to a type $I$ representation of $\frak A_{R}$.
Suppose further that  $\varphi$ is $U(1)$ gauge invariant , $\varphi\circ\gamma_{\theta}=\varphi$.
Then,  $\varphi$ is a product state.
\end{theorem}
\begin{pro}
\label{pro:U(1)2}
Let $\psi$ be a translationally invariant pure state of $\frak A^{CAR}$. 
\\
(i) Suppose further that  $\psi$ is $U(1)$ gauge invariant, $\psi\circ\gamma_{\theta}=\psi$.
The $\Theta$ invariant extension of $\psi_{+}$ to $\frak A$ is a translationally invariant pure
state.
\\
(ii) Suppose further that  $\psi$ is $U(1)$ gauge invariant and the von Neumann algebra
 $\pi_{\psi}(\frak A^{CAR}_{L})^{\dprime}$ associated with the GNS representation of
 $\psi_{L}$ is of type $I$. 
 Then,  either $\psi = \psi_{F}$  or $\psi = \psi_{AF}$ holds.
\end{pro}

Next we present a proof for Theorem \ref{th:Matsui2001} partly different from the one in \cite{Split}. 
Let $\{ \pi (\frak A) , \Omega, \frak H \}$ be the GNS triple for $\varphi$. 
Suppose that $\frak M_{R}$ is of type $I$.
As $\varphi_{R}$ is $\gamma_{\theta}$ invariant, $\gamma_{\theta}$ is extendible to an
$U(1)$ action on the type $I$ factor $\frak M_{R}$.
As any automorphism of an type $I$ factor is inner, 
there exists a projective unitary representation $U_{R}(\theta)$
in $\frak M_{R}$ satisfying
\begin{equation}
 U_{R}(\theta) \pi (Q) U_{R}(\theta)^{*} = \pi (\gamma_{\theta}(Q)) , \quad Q \in \frak A_{R} .
 \label{eqn:a1}
 \end{equation}
For  $U(1)$ the cocycle is trivial and we may assume that  $ U_{R}(\theta)$
is a representation of $U(1)$.
Similarly we obtain  a representation $ U_{L}(\theta)$ of $U(1)$ in $\frak M_{L}$ satisfying
\begin{equation}
U_{L}(\theta) \pi (Q) U_{L}(\theta)^{*} = \pi (\gamma_{\theta}(Q)) ,  \quad Q \in \frak A_{L} .
 \label{eqn:a2}
 \end{equation}
Furthermore by suitably choosing phase factors and setting $U(\theta) = U_{R}(\theta) U_{L}(\theta)$,  
we obtain
\begin{equation}
 U (\theta) \Omega = \Omega ,  \quad U_{R}(\theta)\Omega = U_{L}(-\theta)\Omega
 \label{eqn:a3}
 \end{equation}
We write the Fourrier series for $U_{R}(\theta)$ and $U_{L}(\theta)$ 
as follows:
$$ U_{R}(\theta) = \sum_{k=-\infty}^{\infty}   e^{i k \theta} P_{R}(k), \quad
U_{L}(\theta) = \sum_{k=-\infty}^{\infty}   e^{i k \theta} P_{L}(k) $$ 
Due to (\ref{eqn:a3}) we have
$ P_{R}(k)\Omega = P_{L}(-k)\Omega$.
\bigskip
\noindent
\par
The state $\varphi$ is translationally invariant,  $\tau_{1}$ restricted to $\frak A_{R}$
 is extendible to  the von Neumann algebra $\frak M_{R}$ as an endomorphism denoted by
 $\Xi_{R}$. 
 $$ \Xi_{R} (\pi (Q) ) = \pi ( \tau_{1}(Q)) , \quad Q \in \frak A_{R} .$$
 This endomorphism  $\Xi_{R}$ is a shift of the type $I$ von Neumann algebra 
 $\frak M_{R}$, namely,
$\cap_{k=0}^{\infty}  \Xi^{k} (\frak M_{R})  =   {\bfC} 1$.
Then, there exists a representation of $O_{2}$ in   $\frak M_{R}$  implementing 
$\Xi_{R}$ .
\begin{equation}
T_k^* T_l \: = \: \delta_{kl} 1, \quad
\Xi_{R} (Q) = T_{1} Q T_{1}^{*} +  T_{2} Q T_{2}^{*}  , \quad  Q \in \frak M_{R}
\label{eqn:a4}
 \end{equation}
We can introduce a backward shift  $\Xi_{L}$ on  $\frak M_{L}$ 
satisfying
$$ \Xi_{L} (\pi (Q) ) = \pi ( \tau_{-1}(Q)) , \quad Q \in \frak A_{L}$$
and another representation of $O_{2}$ in   $\frak M_{L}$  implementing 
$\Xi_{L}$ .
\begin{equation}
S_k^* S_l \: = \: \delta_{kl} 1, \quad
\Xi_{L} (Q) = S_{1} Q S_{1}^{*} +  S_{2} Q S_{2}^{*}  , \quad  Q \in \frak M_{L}
\label{eqn:a5}
 \end{equation}
The representations of $O_{2}$ satisfying (\ref{eqn:a4}) and (\ref{eqn:a5}) 
is not unique because we have freedom of the $U(2)$ gauge action 
(or choice of base of the 2 dimensional space) but we may assume that 
$$T_2 T_2^{*} -  T_1 T_1^{*} = \sigma_{z}^{(1)}  , \quad 
S_2 S_2^{*} -  S_1 S_1^{*} = \sigma_{z}^{(0)} .$$
Still we have freedom to choose the phase factor corresponding to the U(1) gauge action. 
If we set $V =  S^{*}_{1}T_{1} +S^{*}_{2}T_{2} $, a direct computation shows
that $V$ is a unitary and
\begin{equation}
 V \pi (Q) V^{*} = \pi (\tau_{1}(Q)) 
\label{eqn:a6}
 \end{equation}
for any $Q$ in $\frak A$.
As the state $\varphi$ is translationally invariant we may assume that
\begin{equation}
V\Omega = \Omega ,  \quad S^{*}_{k}\Omega = T^{*}_{k}\Omega . 
\label{eqn:a7}
 \end{equation}
Next turn to $U_{R}(\theta ) T_{k}U_{R}(\theta )^{*}$. These operators satisfy the relation
of the generators of $O_{2}$. On the other hand, the adjoint action of $U_{R}(\theta ) $
is same as $\gamma_{\theta}$ restricted on $\frak M_{R}$. By this fact we conclude
\begin{equation} 
U_{R}(\theta ) T_{1}U_{R}(\theta )^{*}  = e^{il\theta} T_{1} ,  \quad
U_{R}(\theta ) T_{2} U_{R}(\theta )^{*} = e^{i(l+1)\theta} T_{2} 
\label{eqn:a8}
 \end{equation}
By the same reason,
\begin{equation} 
U_{L}(\theta ) S_{1}U_{R}(\theta )^{*}  = e^{il^{\prime}\theta} S_{1} ,  \quad
U_{R}(\theta ) S_{2} U_{R}(\theta )^{*} = e^{i(l^{\prime}+1)\theta} S_{2} 
\label{eqn:a9}
 \end{equation}
We claim that $l = l^{\prime}$. As $\tau_{1}$ commutes with $\gamma_{\theta}$
$V$ commutes with $U(\theta)$ where we used
$ V\Omega =\Omega ,  U(\theta)\Omega =\Omega$. 
By definition,
$$U(\theta) V =     e^{i(l-l^{\prime})\theta}    VU(\theta)$$
so we conclude $l = l^{\prime}$.
\\
(\ref{eqn:a8}) and (\ref{eqn:a9}) tell us
\begin{eqnarray} 
&&P_{R}(k) T_{1} = T_{1} P_{R}(k-l) , \:\: P_{R}(k) T_{2} = T_{2} P_{R}(k-l-1),
\nonumber
\\
&&P_{L}(k) S_{1} = S_{1} P_{L}(k-l) , \:\: P_{L}(k) S_{2} = S_{2} P_{L}(k-l-1).
\label{eqn:a10}
 \end{eqnarray}
Setting  $S_{1}S_{1}^{*}= e_{1}^{(0)}$,  $S_{2}S_{2}^{*}= e_{2}^{(0)}$
$T_{1}T_{1}^{*}= e_{1}^{(1)}$,  $T_{2}T_{2}^{*}= e_{2}^{(1)}$,
we have
$$(\Omega ,    e_{1}^{(0)} P_{R}(k) \Omega) = (S_{1}^{*} \Omega , P_{R}(k)S_{1}^{*} \Omega) =
(\Omega ,   T_{1} P_{R}(k)T_{1}^{*} \Omega) =(\Omega , P_{R}(k+l) e_{1}^{(1)}\Omega) $$
and 
$$(\Omega ,    e_{2}^{(0)} P_{R}\Omega) = (\Omega ,   P_{R}(k+l+1) e_{2}^{(1)} \Omega) $$
where we used \ref{eqn:a7}.
As $ e_{1}^{(0)} + e_{2}^{(0)} =1= e_{1}^{(1)}+ e_{2}^{(1)}$
\begin{eqnarray}
&&(\Omega ,   P_{R}(k)\Omega )=
(\Omega ,   ( e_{1}^{(1)}+  e_{2}^{(1)})P_{R}(k)\Omega )
\nonumber
\\
&&=(\Omega , P_{R}(k+l) e_{1}^{(1)} +   P_{R}(k+l+1) e_{2}^{(1)}) \Omega) 
\label{eqn:a11}
\end{eqnarray}
Suppose that $l=0$. Then,
$$(\Omega ,    e_{2}^{(1)}P_{R}(k)\Omega )
=(\Omega ,   P_{R}(k+1) e_{2}^{(1)} \Omega) =\alpha $$
for any $k$. Thus, for any $m$, we obtain  
$$(\Omega ,    e_{2}^{(1)} \Omega ) \geq
\sum_{k=n}^{n+m} (\Omega ,   P_{R}(k+1) e_{2}^{(1)} \Omega) = m \alpha .$$
This shows that $\alpha =0$ and 
$$(\Omega ,    e_{2}^{(1)} \Omega ) =
\sum_{k=-\infty}^{\infty} (\Omega ,   P_{R}(k+1) e_{2}^{(1)} \Omega)=0$$
Thus , $\varphi$ is a translational invariant pure state satisfying
$\varphi (e_{1}^{(1)})=0 $  which is a product state.
\\
Suppose that $l=-1$. Then,
$$(\Omega ,    e_{1}^{(1)}P_{R}(k)\Omega )
=(\Omega ,   P_{R}(k-1) e_{1}^{(1)} \Omega) =\alpha $$
for any $k$. By the same line of reasoning
$$(\Omega ,    e_{1}^{(1)} \Omega ) =0$$
Thus , $\varphi$ is a translational invariant pure state satisfying
$\varphi (e_{2}^{(1)})=0 $  which is a product state.
\\
Suppose that $l\geq 1$. Take sum of $k$ in (\ref{eqn:a11})
$$\sum_{k=n}^{\infty} (\Omega ,   ( e_{1}^{(1)}+  e_{2}^{(1)})P_{R}(k)\Omega )
=\sum_{k=n}^{\infty} (\Omega , P_{R}(k+l) e_{1}^{(1)} + P_{R}(k+l+1) e_{2}^{(1)}) \Omega)$$ 
It turns out
\begin{equation}
\sum_{k=n}^{l-1} (\Omega ,   e_{1}^{(1)}P_{R}(k)\Omega )
+\sum_{k=n}^{l} (\Omega ,     e_{2}^{(1)}P_{R}(k)\Omega ) =0
\label{eqn:a12}
\end{equation}
Each summand is positive in (\ref{eqn:a12}) and we see
$$ (\Omega ,   e_{1}^{(1)}P_{R}(k)\Omega )= (\Omega ,     e_{2}^{(1)}P_{R}(k)\Omega ) =0$$
This shows $ (\Omega ,   e_{1}^{(1)}\Omega )=0$ $ (\Omega ,     e_{2}^{(1)} \Omega ) =0$
and we arrive at a contradiction. So $l\geq 1$ is not possible.
Similarly $l\leq -2$ is impossible.
{\it End of Proof}
\bigskip
\bigskip
\noindent
\newline
 {\it Proof of Proposition \ref{pro:U(1)2}}
\\
To prove Proposition \ref{pro:U(1)2} (i), we show
the case (iii) in Proposition \ref{pro:ext1}  is impossible due to assumption of 
$\gamma_{\theta}$ invariance.
There exists $U(\theta)$ implementing $\gamma_{\theta}$ 
on the GNS space of $\psi$. Then
$$U(\theta)  V(\Theta_{-}) U(\theta)^{*} = c(\theta) V(\Theta_{-}) $$
as the adjoint action of both unitaries are identical. Moreover these are selfadjoint 
so   $c(\theta)=\pm 1$ . Due to continuity in $\theta$ we conclude that $c(\theta)=1$
and $V(\Theta_{-})$ is an even element.
\par
Finally, we consider Proposition \ref{pro:U(1)2} (ii).  Due to (i) of Proposition \ref{pro:U(1)2} (i),
the Fermionic state $\psi$ has a translationally invariant pure state extension $\varphi$ to $\frak A$.
Then, the split property for Fermion implies that that of the Pauli spin system. It turns out that 
either $\psi (c^{*}_{j}c_{j}) =\varphi (e^{(j)}_{1}) = 0 $
or $\psi (c_{j}c_{j^{*}}) =\varphi (e^{(j)}_{2}) = 0$ holds. This completes our proof of Proposition 
\ref{pro:U(1)2} (ii). 
\\
 {\it End of Proof of Proposition \ref{pro:U(1)2}}


\end{document}